%% file: hqet_spectrum.tex
\documentclass[11pt]{article}
\usepackage{amsmath,amsfonts,bbm,sint,epsfig,cite,graphics,mathrsfs}
\usepackage{macros_static}
\usepackage{wrapfig}

\begin{document}

\input title.tex

\input s1.tex

\input s2.tex

\input s3.tex

\input s4.tex
\vskip2ex\noindent
{\bf Acknowledgements.}
We thank Jochen Heitger and Patrick Fritzsch for useful communications.

This work is supported
by the Deutsche Forschungsgemeinschaft
in the SFB/TR~09,
and by the European community
through EU Contract No.~MRTN-CT-2006-035482, ``FLAVIAnet''.
N.G. acknowledges financial support from the MICINN grant FPA2006-05807,
the Comunidad Aut\'onoma de Madrid programme HEPHACOS~P-ESP-00346,
and participates in the Consolider-Ingenio 2010~CPAN (CSD2007-00042).
T.M. thanks the A.~von~Humboldt~Foundation for support.

\begin{appendix}
\input a1.tex

\end{appendix}

\bibliographystyle{JHEP}
\bibliography{hqet_spectrum}

\end{document}

%% file: title.tex
\begin{titlepage}

\begin{flushright}
\vskip 0.7cm
DESY 09-154 \\
SFB/CPP-10-02\\
Edinburgh 2010/07\\
MKPH-T-10-05\\
LPT-Orsay/10-09\\
\end{flushright}

\vskip 0.35cm
\begin{center}
{\Large\bf 
HQET at order $1/m$: II. Spectroscopy in the quenched approximation
}
\end{center}
\vskip 0.35cm
\vbox{
\centerline{
\epsfxsize=2.8 true cm
\epsfbox{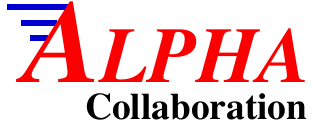}}
}
\vskip 0.1cm
\begin{center}
{
Beno\^it~Blossier$^{\scriptscriptstyle a}$,
Michele~Della~Morte$^{\scriptscriptstyle b}$,
Nicolas~Garron$^{\scriptscriptstyle c,d}$,
Georg~von~Hippel$^{\scriptscriptstyle b,e}$,
Tereza~Mendes$^{\scriptscriptstyle e,f}$,
Hubert~Simma$^{\scriptscriptstyle e}$
and Rainer~Sommer$^{\scriptscriptstyle e}$ 
}
\vskip 0.5cm
{
\vskip 2.0ex
$^{\scriptstyle a}$
Laboratoire de Physique Th\'eorique,
B\^atiment 210, Universit\'e Paris XI,
F-91405~Orsay~Cedex, France
\vskip 2.0ex
$^{\scriptstyle b}$
Institut f{\"u}r Kernphysik, University of Mainz,
D-55099~Mainz, Germany
\vskip 2.0ex
$^{\scriptstyle c}$
Instituto de F\'isica Te\'orica, Universidad Aut\'onoma de Madrid,
Campus~de~Cantoblanco, C/~Fco.~Tom\'as~y~Valiente~7, 
Madrid~E-28049, Spain
\vskip 2.0ex
$^{\scriptstyle d}$
SUPA, School of Physics, University of Edinburgh,
Edinburgh EH9 3JZ, U.K.
\vskip 2.0ex
$^{\scriptstyle e}$
NIC, DESY,
Platanenallee 6, 15738~Zeuthen, Germany
\vskip 2.0ex
$^{\scriptstyle f}$
IFSC, University of S\~ao Paulo, 
C.P.~369, CEP~13560-970, S\~ao~Carlos~SP, Brazil
}
\vskip 0.775cm
{\bf Abstract}
\vskip 0.1ex
\end{center}
Using Heavy Quark Effective Theory
with non-perturbatively determined parameters
in a quenched lattice calculation,
we evaluate the splittings between the ground state 
and the first two radially excited states
of the $B_s$ system at static order.
We also determine the splitting between
first excited and ground state,
and between the $B_s^*$ and $B_s$ ground states
to order $\minv$.
The Generalized Eigenvalue Problem and
the use of all-to-all propagators
are important ingredients of our approach.

\enlargethispage{2ex}
\vskip 2.0ex
\noindent{\it Key words:}
Lattice QCD; Heavy Quark Effective Theory; Hadronic Energy Levels
\vskip 2.0ex
\noindent{\it PACS:}
12.38.Gc; %Lattice QCD calculations
12.39.Hg; %Heavy quark effective theory
14.40.Nd  %Bottom mesons

\end{titlepage}

%% file: s1.tex
\section{Introduction}

In recent years, there has been significant progress in determining
the spectrum of hadrons containing a $b$ quark,
both experimentally
\cite{D0:2007sna, %B_s^* excitation by D0
D0:2007vq, %B and B^* excitations by D0
CDF:2008jn, %B excitations by CDF
CDF:2007tr, %B_s excitations by CDF
Taylor:2009zzb},
and on the lattice
\cite{Wingate:2002fh, %HPQCD
Foley:2007ui, %Dublin
Koponen:2007nr, %UKQCD
Jansen:2008si, %ETMC
Burch:2008qx}. %BGR
A comparison of theory and experiment 
is of considerable interest for these hydrogen like systems,
in particular since
Heavy Quark Effective Theory (HQET)
\cite{stat:eichhill1,stat:symm1,stat:symm3,Eichten:1990vp}
is applicable and is at the same time an important theoretical tool
to isolate new physics in the flavor sector 
\cite{CKM08}.

The extraction of information on excited states from lattice simulations
is a difficult problem, since the excited states appear as subleading
exponentially decaying contributions to the lattice correlators, which
at large times are dominated by the ground state and at small times
by the combined contributions of arbitrarily highly excited states,
and on top of this are affected by noise. A number of different methods
\cite{gevp:michael,phaseshifts:LW, %variational method
Lepage:priors, %Bayesian fitting
vonHippel:2007ar} %Evolutionary fitting
have been proposed for overcoming this challenge;
recently we have shown that with an efficient use of the generalized
eigenvalue problem (GEVP), rigorous statements about the systematic
error caused by the admixture of other states can be proven
\cite{gevp:pap}.
In particular, we have shown that for a suitable choice of Euclidean times
$t$ and $t_0$, the systematic error decays exponentially with
an exponent given by an energy gap that can be made large by an
appropriately chosen variational basis.

Since their Compton wavelength is much shorter than any realistically
achievable lattice spacing, $b$ quarks cannot be simulated as relativistic
quarks on a lattice.
A description of heavy-light mesons that is suitable for use in the
context of lattice QCD simulations is given by
HQET with non-perturbatively determined parameters
\cite{Heitger:2003nj,DellaMorte:2006cb}.
The parameters necessary to match HQET to QCD at order
$\rmO(\minv)$ have been determined
in the quenched approximation by our collaboration
\cite{hqet:first1},
and the non-perturbative determination for QCD with $N_f=2$ dynamical
quark flavors is far advanced
\cite{lat08:patrick}.

In this paper, we present our results for the heavy-light meson
spectrum from HQET using the GEVP method. In section
\ref{s:methods}, we give a brief review of our methods. Details of
the simulations and data analysis procedures are given in section
\ref{s:details}, and our results for the quenched heavy-light spectrum
are presented. Section \ref{s:conclusions} contains our conclusions.
Some technical details of our implementation of all-to-all propagators
are relegated to the appendix.

%% file: s2.tex
\section{Methodological Background}
\label{s:methods}

\subsection{Non-perturbative HQET}

HQET on the lattice offers a theoretically rigorous approach to the
physics of B-mesons since it is based on a strict expansion of QCD
correlation functions in powers of $\minv$ around the limit
$m_{\rm b}\to\infty $.
Subleading effects are described by insertions of higher dimensional
operators whose coupling constants are formally $\rmO(\minv)$
to the appropriate power. This means that HQET can be renormalized
and matched to QCD in a completely non-perturbative way
\cite{lat02:rainer},
implying the existence of the continuum limit at any fixed order
in the $\minv$ expansion.

To fix the notation we write the HQET action at $\rmO(\minv)$ as
\bes\label{e:hqetaction}
S_{\rm HQET} 
&=& a^4\sum_x \left\{ \mathscr{L}_{\rm stat}(x) 
    - \omega_{\rm kin} \Okin(x) 
    - \omega_{\rm spin} \Ospin(x) \right\} \,, 
\\
\label{e:lstat}
\mathscr{L}_{\rm stat}(x) &=& 
\heavyb(x) \left(D_0 + \dmstat \right)\psi_{\rm h}(x)\,, \\
\label{e:ofirst}
\Okin(x) &=& \heavyb(x) \vecD^2\psi_{\rm h}(x)\,,\quad 
\Ospin(x) \;=\; \heavyb(x) \vecsigma \cdot \vecB \psi_{\rm h}(x)\,,
\ees
where $\psi_{\rm h}$ satisfies $\frac{1+\gamma_0}{2}\psi_{\rm h}=\psi_{\rm h}$.
The parameters $\omega_{\rm kin}$ and $\omega_{\rm spin}$ are 
of order $\minv$, and $\dmstat$ is the counter-term 
absorbing the power-divergences of the static quark 
self energy. The signal-to-noise ratio 
of large-distance correlation functions is significantly
improved by replacing the link $U(x,0)$ in the backward covariant derivative
$D_0 f(x) = [f(x) - U^\dagger(x-a\hat0,0)f(x-a\hat0)]/a$,  with a smeared link
\cite{DellaMorte:2005yc}.

Exponentiating the action of \eq{e:hqetaction} would give (non-renormalizable)
NRQCD
\cite{nrqcd:first};
in order to retain the renormalizability of the static theory,
we treat the theory in a strict expansion in $\minv$, where the $\rmO(\minv)$
parts of the action appear as insertions in correlation functions.
For the expectation value of some operator $O$ this means (ignoring the
possibility of explicit $\minv$ operator corrections, which do not affect
the energy levels \cite{hqet:first1})
\be
\langle O\rangle = \langle O\rangle_{\rm stat} +
\omega_{\rm kin}\,a^4\sum_x \langle O\,\Okin(x)\rangle_{\rm stat} +
\omega_{\rm spin}\,a^4\sum_x \langle O\,\Ospin(x)\rangle_{\rm stat}
\ee
where $\langle O\rangle_{\rm stat}$ denotes the expectation value in the static approximation.

To fully specify HQET, the parameters 
$\delta m$, $\omega_{\rm kin}$ and $\omega_{\rm spin}$
must be determined by matching to QCD.
In order to retain the asymptotic convergence in $\minv$
this matching must be done non-perturbatively, since for a
perturbative matching at loop order $l$,
the O($g^{2l}$) truncation error of the static term is 
much larger than the power corrections to the static limit:
\be
\epsilon_{\rm pert}(l) \propto \bar{g}^{2l}\left(m_{\rm b}\right)
\propto \left[2b_0\log(m_{\rm b}/\Lambda_{\rm QCD})\right]^{-l}
\mathop{\gg}\limits_{m_{\rm b}\to\infty} {\Lambda_{\rm QCD} \over m_{\rm b}} \,.
\ee

A fully non-perturbative
determination of the parameters of HQET has been carried out in
\cite{hqet:first1}. 
Here we employ the same discretization
of QCD and HQET and in particular the determined values of $\omega_{\rm kin}$ and
$\omega_{\rm spin}$.
For further details of the matching and discretization, the reader is
referred to
\cite{hqet:first1}.

\subsection{The Generalized Eigenvalue Problem}
\def\first{\mrm{x}}
\def\stat{\mrm{stat}}

In this section, we recall the relevant contents of 
\cite{gevp:pap}.
Starting from some fields $O_i(x)$ localised on a time slice and their 
momentum zero projection $a^3\sum_\vecx O_i(x)=\tilde O_i(x_0)$,
a matrix of Euclidean space correlation functions, 
\bes
  \label{e:cij}
  C_{ij}(t) &=& \langle \tilde O_i(t) \tilde O_j^*(0) \rangle =
  \sum_{n=1}^\infty \rme^{-E_n t} \psi_{ni}\psi_{nj}^*\,,\quad
  i,j = 1,\ldots,N  \\ && \quad \psi_{ni} \equiv (\psi_n)_i =
  \langle 0|\hat O_i|n\rangle
  \quad
  E_n < E_{n+1} \,,
 \nonumber
\ees
provides the basis for the GEVP 
\bes \label{e:gevp}
  C(t)\, v_n(t,t_0) = \lambda_n(t,t_0)\, C(t_0)\,v_n(t,t_0) \,,
  \quad n=1,\ldots,N\,,\quad t>t_0\,.
\ees
Effective energies for the $n$-th energy level are given by
\be
  \label{e:eneff1}
  E_n^\mrm{eff}(t,t_0) = -\partial_t \log\lambda_n(t,t_0) \equiv
    -{1\over a} \, [\log{\lambda_n(t+a,t_0)}-\log{\lambda_n(t,t_0)}]\,,
\ee
where $a$ is the lattice spacing. Provided that $t_0>t/2$,
the effective energies converge to the exact energy levels as
\cite{gevp:pap}
\be
  \label{e:eneff2}
  E_n^\mrm{eff}(t,t_0) =  E_n + \rmO(\rme^{-\Delta E_{N+1,n}\, t}) \,,\quad
                   \Delta E_{m,n} = E_m-E_n.
\ee

In HQET, all correlation functions
\bes
  C_{ij}(t) &=& C_{ij}^{\stat}(t) \,+\,
  \omega_{\rm kin} \,C_{ij}^{\rm kin}(t) \,+\,
  \omega_{\rm spin} \,C_{ij}^{\rm spin}(t) \,+\, \rmO(\omega^2)
\ees
are computed in an expansion in a small parameter, $\omega\propto\minv$.
Correspondingly, the energy levels expand as
\bes
    E_n^{\rm eff}(t,t_0) &=& E_n^{\rm eff,\stat}(t,t_0)
     +\omega_\first E_n^{\rm eff,\first}(t,t_0) +\rmO(\omega^2)\\
    E_n^{\rm eff,\stat}(t,t_0) &=& a^{-1}\,\log {\lambda_n^\stat(t,t_0) \over
                         \lambda_n^\stat(t+a,t_0)}
   \label{e:lamstat}
 \\[1mm]
     E_n^{\rm eff,\first}(t,t_0) &=& {\lambda_n^\first(t,t_0) \over
                         \lambda_n^\stat(t,t_0)} \,-\,
      {\lambda_n^\first(t+a,t_0) \over
                         \lambda_n^\stat(t+a,t_0)}
   \label{e:lamfirst}
\ees
where $\first\in\{\rm kin,\rm spin\}$,
with the behavior at large time $t\le 2t_0$,
\bes
    E_n^{\rm eff,\stat}(t,t_0) &=&
    E_n^\stat \,+\, \beta_n^\stat\,\rme^{-\Delta E_{N+1,n}^\stat\, t}+\ldots\,,
   \label{e:lamstatfit}
  \\
    E_n^{\rm eff,\first}(t,t_0) &=&
    E_n^\first \,+\, [\,\beta_n^\first
               \,-\, \beta_n^\stat\,t\,\Delta E_{N+1,n}^\first\,]
                     \rme^{-\Delta E_{N+1,n}^\stat\, t}+\ldots\, .
   \label{e:lamfirstfit}
\ees
where
\bes
    C^\stat(t) \,v_n^\stat(t,t_0) &=& \lambda_n^\stat(t,t_0)\, C^\stat(t_0)
             \,v_n^\stat(t,t_0) \,,\\
  {\lambda_n^\first(t,t_0)\over \lambda_n^\stat(t,t_0)}
  &=& \left(v_n^\stat(t,t_0)\,,\,
             [[\lambda_n^\stat(t,t_0)]^{-1}\,C^{\first}(t)- C^{\first}(t_0)]
             v_n^\stat(t,t_0)\right).\nonumber
\ees
with 
$
          (v_m^\stat(t,t_0)\,,\,C^\stat(t_0)\,v_n^\stat(t,t_0)) =\delta_{mn}.
$
We note that the GEVP is only ever solved for the static correlator
matrices.

\subsection{Mass splittings to $\rmO(\minv)$}

With the static Lagrangian \eq{e:lstat}, all HQET energies satisfy exactly
$E_n = \left.E_n\right|_{\dmstat=0} +{1\over a} \log(1+a\dmstat)$, and
the power divergent $\dmstat$ drops out in energy differences.
Since we only consider these in this paper, we need just
$\omegakin$ and $\omegaspin$ of \cite{hqet:first1}.

The excitation energies at the static order of HQET are given simply
by the differences of the static energies,
\be
\Delta E_{n,1}^\stat = E_n^\stat - E_1^\stat \,,
\ee
and at order $\minv$, the excitation energies of pseudoscalar
$B_s$ states become
\be
\Delta E_{n,1}^{\rm HQET} = (E_n^\stat - E_1^\stat) + 
                       \omega_{\rm kin} (E_n^{\rm kin} - E_1^{\rm kin}) +
                       \omega_{\rm spin} (E_n^{\rm spin} - E_1^{\rm spin})\,.
\ee

At the static order, the masses of pseudoscalar and vector states are
degenerate due to spin symmetry
\cite{stat:symm3}.
This degeneracy is lifted at the $\rmO(\minv)$ level by the contribution
from $\Ospin$, giving a $B_s-B_s^*$ mass difference of
\be
\Delta E_{\rm P-V} = \frac{4}{3} \omega_{\rm spin} E_1^{\rm spin} \,.
\ee

%% file: s3.tex
\section{Simulation details and results}
\label{s:details}

\subsection{Lattice parameters}

We use three quenched ensembles of 100 configurations each,
generated using the Wilson gauge action with parameters
$\beta=6.0219$, $6.2885$ and $6.4956$. The physical
volume was kept constant at $L\approx 1.5$ fm, leading to
$L/a=16$, $24$, $32$ for the three ensembles.
We used time extent $T=2L$ throughout.

The static quark is discretized on each ensemble with both
the HYP1 and HYP2
\cite{HYP,HYP:pot,DellaMorte:2005yc}
actions.
The light valence quark is discretized by a non-perturbatively
$\rmO(a)$-improved Wilson action
\cite{impr:SW,impr:pap3},
and its mass was fixed to the strange quark mass,
giving $\kappa_s=0.133849$, $0.1349798$, $0.1350299$, respectively
\cite{mbar:pap3}.
A summary of the simulation parameters used, with the corresponding
values of $\omegakin$ and $\omegaspin$, is given
in \tab{tab:parameters}.

\begin{table}
\begin{center}
\begin{tabular}{lrllrrll}\hline\hline
$\beta$&$r_0/a$& $(L/a)^3\times T/a$    & $\kappa_s$&$N_L$&$N_\eta$&
$\omegakin/a$&$\omegaspin/a$\\\hline
6.0219 & 5.6  & $16^3 \times 32$ & 0.133849  & 50 & 2 & 0.330(5) & 0.55(4) \\
6.2885 & 8.4  & $24^3 \times 48$ & 0.1349798 & 50 & 2 & 0.439(7) & 0.71(5) \\
6.4956 & 11.0 & $32^3 \times 64$ & 0.1350299 &  0 & 4 & 0.553(9) & 0.87(6) \\\hline\hline
\end{tabular}
\end{center}
\caption{Parameters of the simulations: inverse coupling $\beta$,
approximate value of the scale parameter $r_0$
\protect\cite{pot:r0} 
in lattice units, spacetime volume,
hopping parameter for the strange quark mass
\protect\cite{mbar:pap3},
number of low-lying eigenmodes and number of noise sources
used in the all-to-all
\protect\cite{alltoall:dublin}
estimate of the strange quark propagators,
and values of the HQET couplings $\omegakin$
and $\omegaspin$
\protect\cite{hqet:first1}
for the HYP2 action in lattice units.}
\label{tab:parameters}
\end{table}

\subsection{Measurements of correlation functions}

The strange quark propagators are computed through a variant of
the Dublin all-to-all method
\cite{alltoall:dublin}.
We use approximate instead of exact low modes (the method remains exact)
and employ even-odd preconditioning in order to reduce the size of the
stochastically estimated inverse of the Dirac operator by a factor of 2.
The reader is referred to appendix 
\ref{app:ata} 
for details.

The interpolating fields are constructed using quark bilinears
\begin{eqnarray}
O_k(x) &=& \psibar_{\rm h}(x)\gamma_0\gamma_5\psi_{\rm l}^{(k)}(x) \\
O_k^*(x) &=& \psibar_{\rm l}^{(k)}(x)\gamma_0\gamma_5\psi_{\rm h}(x) \nonumber
\end{eqnarray}
built from the static quark field $\psi_{\rm h}(x)$ and different levels
of Gaussian smearing
\cite{wavef:wupp1}
for the light quark field
\begin{equation}
\psi_{\rm l}^{(k)}(x) = \left( 1+\kappa_{\rm G}\,a^2\,\Delta \right)^{R_k} \psi_{\rm l}(x)\,,
\end{equation}
where the gauge fields in the covariant Laplacian $\Delta$ are first
smeared with 3 iterations of (spatial) APE smearing
\cite{smear:ape,Basak:2005gi}
to reduce noise.
At $\beta=6.2885$, we use $R_k=0$, $22$, $45$, $67$, $90$, $135$, $180$, $225$
with $\kappa_{\rm G}=0.1$. At the other values of $\beta$, we rescale the values of
$R_k$ used so that the physical size
$r_{{\rm phys},k}\approx 2 a \sqrt{\kappa_{\rm G} R_k}$
of the wavefunctions is kept fixed; in particular we keep
$r_{\rm max}=r_{{\rm phys},7}\approx 0.6$ fm.

For these bilinears, we compute the following correlators:
\bea
\nonumber
C^{\rm{stat}}_{ij}(t)& = & \sum_{x,\bf{y}}
\left< O_i(x_0+t,{\bf y})\,O^*_j(x)\right>_\stat,\\[-0.5ex]
\label{e:cmatdefs}
\\[-0.5ex]
\nonumber
C^{\rm{kin/spin}}_{ij}(t)& = & \sum_{x,\vecy,z}\;
\left< O_i(x_0+t,\vecy)\,O^*_j(x)\,  {\cal O}_{\rm kin/spin}(z)\right>_\stat
\eea
with the $\rmO(\minv)$ fields defined in \eq{e:ofirst}.

\subsection{Determination of energies}

The correlator matrices of \eq{e:cmatdefs} are ``thinned'' to form
a sequence of $N\times N$ matrices, where $N\in\{2,\dots,7\}$, by
selecting only entries from a certain subset $\mathcal{I}_N$ of indices
where $\mathcal{I}_N = \{1,7\}$, $\{1,4,7\}$, $\{1,3,5,7\}$, $\{1,2,4,6,7\}$,
$\{1,2,3,5,6,7\}$, $\{1,2,3,4,5,6,7\}$.
An alternative procedure, used in
\cite{gevp:pap}
is ``pruning''
\cite{gevp:bern},
where the $N\times N$ matrices are formed by projection onto
the subspaces spanned by the lowest $N$ eigenvectors of $C(t_i)$
for some small $t_i$. We decided not to use this version since it
introduces a dependence on the relative normalization of the fields
$O_i$. Moreover, with thinning we found a somewhat faster convergence 
to the plateau for
the ground state energy compared to the pruning version 
\cite{lat09:Tereza}
when the normalization of $O_i$ in
\cite{gevp:pap}
is used.
We note that the fact that we found the same low-lying 
energy levels with thinning as with pruning is a further
confirmation that the GEVP
is quite robust against changes of the variational basis employed.

For each of the resulting $N\times N$ correlator matrices,
we solve the static GEVP and compute the static
and $\rmO(1/m_{\rm b})$ energies.
This gives a series of estimates
$E_n^{\rm eff, stat}(N,t,t_0)$,  $E_n^{\rm eff, kin}(N,t,t_0)$ and 
$E_n^{\rm eff, spin}(N,t,t_0)$ with associated statistical errors,
which we determine by a full Jackknife analysis.

To arrive at final numbers for $E_n$ we also need to estimate
the size of the systematic errors coming from the higher
excited states. 
To do this, we first perform a fit of the form
\bea
E_n^{\rm eff,\rm stat}(N,t,t_0) &=& E_n^{\rm stat} + \epsilon_n^{N,\rm stat}(t) \\
 &=& E_n^{\rm stat} + \beta^{\rm stat}_{n,N} \rme^{-(E^{\rm stat}_{N+1}-E^{\rm stat}_n)t}\nonumber
\eea
to the GEVP results for $E_n^{\rm eff,\rm stat}(N,t,t_0)$, fitting the data
at $N=3,\dots,5$, $\frac12 r_0 < t_0 \leq 6a$, $t_0\leq t\leq 2t_0$ and
$n=1,\dots,6$ simultaneously.
The stability of the fit is enhanced in the following manner:
First we perform an unconstrained fit to extract
$E_n^\stat-E_1^\stat$ for $\;n=4,5,6$ for each lattice spacing and action and
compute a rough average of $r_0(E_n^\stat-E_1^\stat)$ for these
values of $n$. Then we repeat the fit, constraining
$r_0(E_n^\stat-E_1^\stat)$ for $n\geq 4$ to the previous average
(this renders the fit linear). 
For $n\leq3$ this is unnecessary as these levels are well determined
at each individual lattice spacing. We list the extracted values of
$r_0(E_n^{\rm stat}-E_1^{\rm stat})$ in \tab{tab:extracted}.

\begin{table}
\begin{center}
\begin{tabular}{|c|cc|ccc|}\hline
$n$                                  &      2 &    3 &  4 &  5 &  6 \\\hline
$r_0(E_n^{\rm stat}-E_1^{\rm stat})$ & 1.50(5)&2.7(1)& 4.0& 5.0& 6.0\\\hline
\end{tabular}
\end{center}
\caption{Rough estimate of the static spectrum. For $n\le 3$, the gaps
come from the continuum limit; for $n\ge 4$, the rough estimate used
to stabilize the fit is quoted.}
\label{tab:extracted}
\end{table}

Finally, using the values of $E_n^{\rm stat}$ and $\beta^{\rm stat}_{n,N}$
determined from this fit as (fixed) input parameters, we fit
$E_n^{\rm eff,\rm kin}(N,t,t_0)$ and $E_n^{\rm eff,\rm spin}(N,t,t_0)$ by
\bea
E_n^{\rm eff,\rm kin}(N,t,t_0) &=& E_n^{\rm kin} + \epsilon_n^{N,\rm kin}(t)\\
 &=& E_n^{\rm kin} + \left[
\beta^{\rm kin}_{n,N} - \beta^{\rm stat}_{n,N}\,t\,(E^{\rm kin}_{N+1}-E^{\rm kin}_n)
\right]\rme^{-(E^{\rm stat}_{N+1}-E^{\rm stat}_n)t} \nonumber \\
E_n^{\rm eff,\rm spin}(N,t,t_0) &=& E_n^{\rm spin} + \epsilon_n^{N,\rm spin}(t)\\
 &=& E_n^{\rm spin} + \left[
\beta^{\rm spin}_{n,N} - \beta^{\rm stat}_{n,N}\,t\,(E^{\rm spin}_{N+1}-E^{\rm spin}_n)
\right]\rme^{-(E^{\rm stat}_{N+1}-E^{\rm stat}_n)t} \nonumber
\eea
in the same manner.

While the fitted results are quite stable, we consider
them as rough estimates only, since our fits include only the
leading exponential correction, and there are systematic
effects from higher corrections to the GEVP.
We therefore employ the fits only to estimate the size of the
leading corrections.
For a reliable estimate of the energy levels,
we calculate plateau averages
from $t=t_{\rm min}\ge t_0$ on at each $N$ and $t_0$,
and take as our final estimate that plateau average for which the
sum $\sigma_{\rm tot}=\sigma_{\rm stat}+\sigma_{\rm sys}$
of the statistical error $\sigma_{\rm stat}$ of the plateau average and
the maximum systematic error $\sigma_{\rm sys}=\epsilon(t_{\rm min})$
becomes minimal,
subject to the constraint that $\sigma_{\rm sys}<\frac13 \sigma_{\rm stat}$.
We impose the latter constraint in order to ensure that the
total error is dominated by statistical errors.

\begin{figure}
\begin{center}
\includegraphics[width=0.45\textwidth]{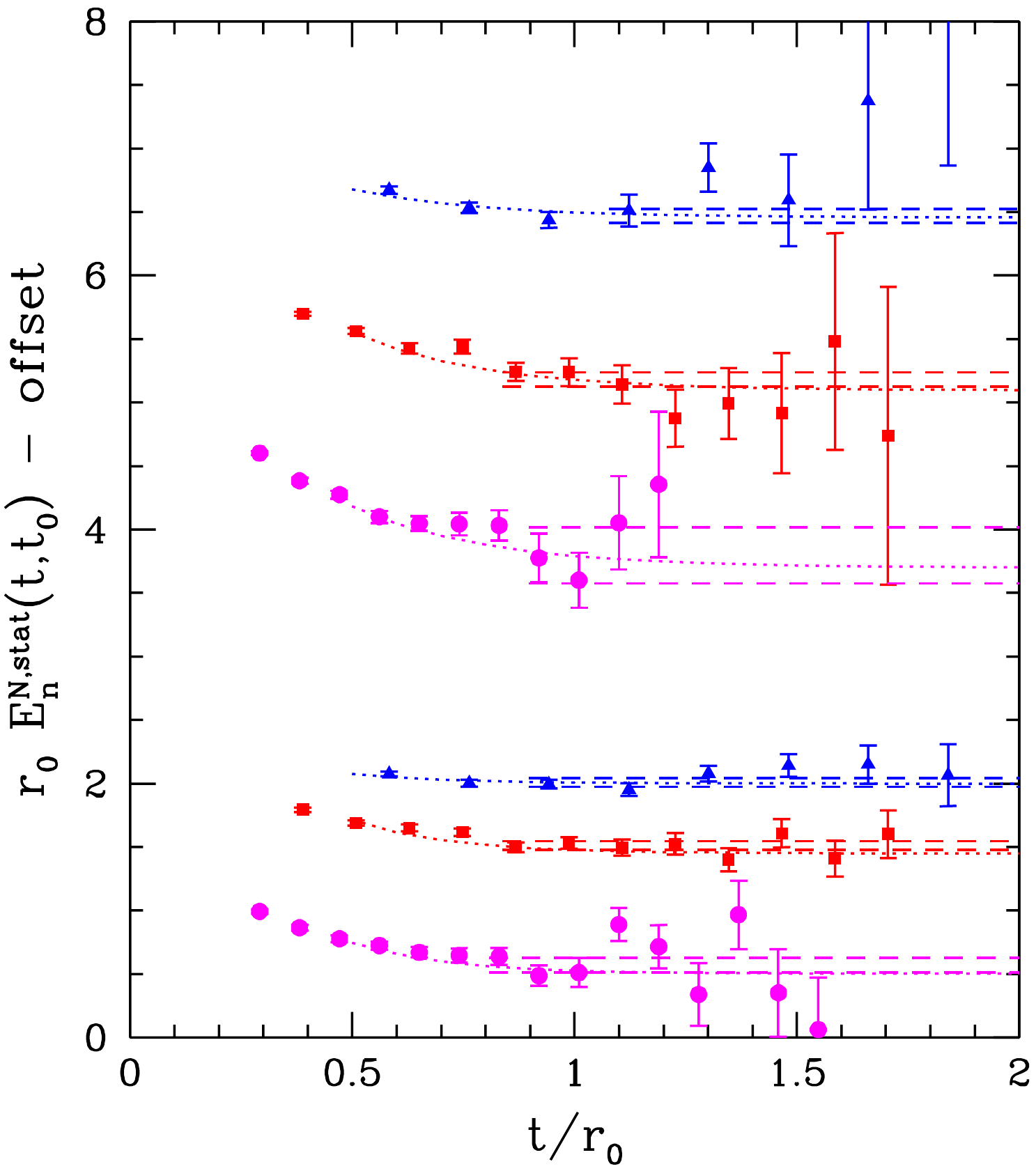}
\includegraphics[width=0.4815\textwidth]{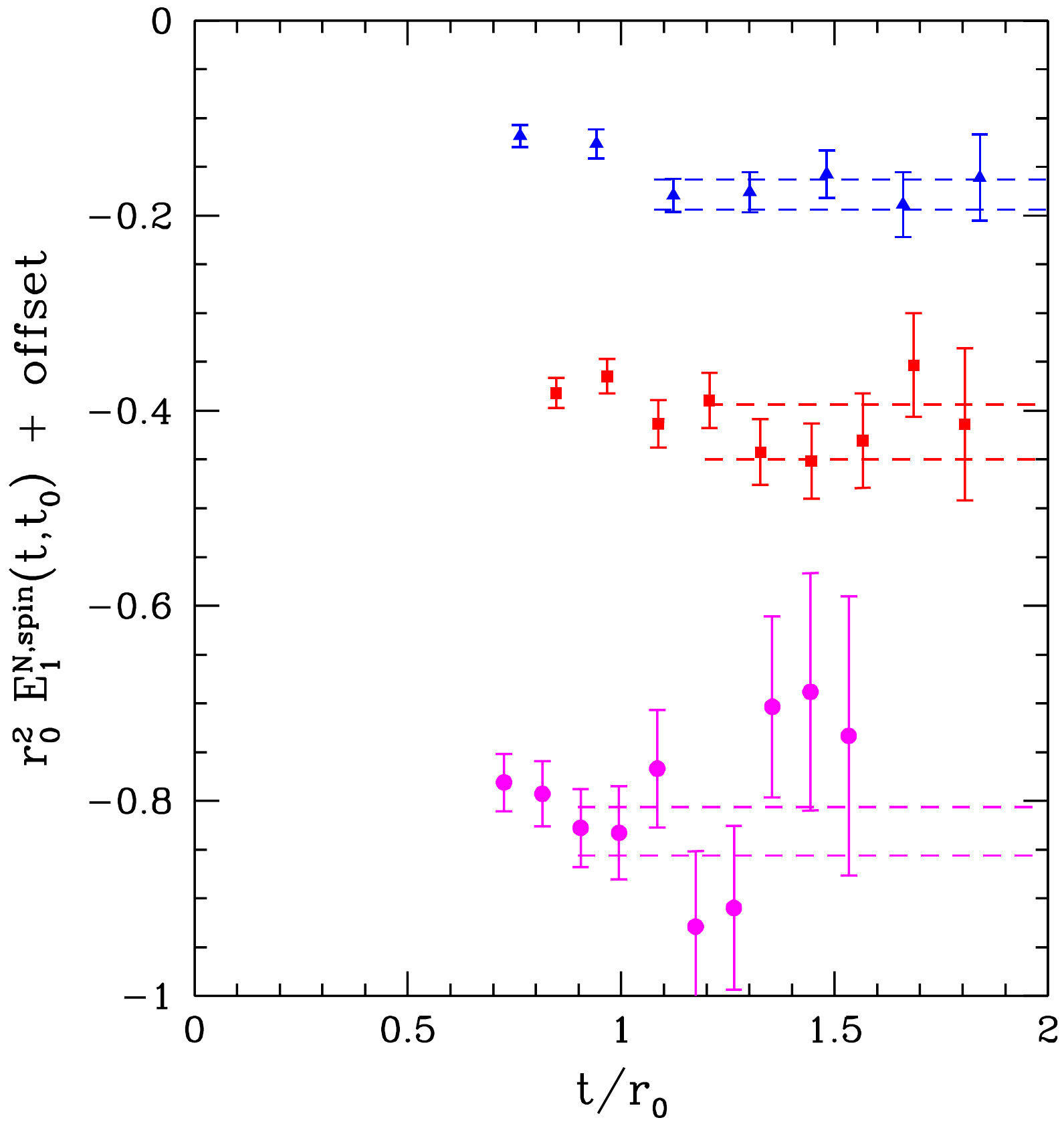}
\end{center}
\caption{Illustration of some plateaux. Left: 
$E_{2}^{N \rm stat}$ (bottom half) and $E_{3}^{N \rm stat}$ (upper half).
Here $N=5$ and within each group the lattice spacing is decreasing
from top to bottom. Dotted lines represent the global fit,
while dashed lines indicate the chosen plateau. Right: 
the ground state spin splitting. The ordinates of the points are shifted
by an arbitrary $\beta$-dependent offset in order to make the different
plateaux visible separately.}
\label{fig:plateaux}
\end{figure}
An illustration of the more problematic
plateaux is shown in \fig{fig:plateaux}. It is rather clear that without 
some analysis of corrections due to excited states it is very difficult
to locate a safe plateau region at least for $n=3$. 

Our results are given in \tab{tab:all_ens}; besides the plateaux
found by the method described in the preceding paragraph, we also show
the results of the global fit, which in almost all cases agrees very well
with the final plateau value.

\begin{table}
\begin{center}
\begin{tabular}{llllll}
\hline\hline
        &            &\multicolumn{2}{c}{HYP1}&\multicolumn{2}{c}{HYP2} \\
$\beta$ & Observable & Fit & Plateau & Fit & Plateau \\\hline
6.0219 & $aE_1^{\rm stat}$ & 0.4407(2) & 0.441(1) & 0.4081(2) & 0.409(1) \\
       & $aE_2^{\rm stat}$ & 0.708(2) & 0.711(5) & 0.678(2) & 0.680(6) \\
       & $aE_3^{\rm stat}$ & 0.893(5) & 0.92(2) & 0.868(5) & 0.89(2) \\
       & $a^2E_1^{\rm kin}$ & 0.743(1) & 0.740(2) & 0.774(1) & 0.771(2) \\
       & $a^2E_2^{\rm kin}$ & 0.843(8) & 0.85(1) & 0.859(7) & 0.88(1) \\
       & $a^2E_1^{\rm spin}$ & -0.0300(5) & -0.0293(7) & -0.0285(4) & -0.0279(8) \\
       & $a^2E_2^{\rm spin}$ & -0.026(2) & -0.029(3) & -0.023(2) & -0.025(2) \\\hline
6.2885 & $aE_1^{\rm stat}$ & 0.3319(2) & 0.3328(7) & 0.3032(2) & 0.3041(7) \\
       & $aE_2^{\rm stat}$ & 0.507(1) & 0.515(4) & 0.478(1) & 0.485(4) \\
       & $aE_3^{\rm stat}$ & 0.648(3) & 0.67(1) & 0.614(3) & 0.625(7) \\
       & $a^2E_1^{\rm kin}$ & 0.6479(5) & 0.6481(4) & 0.6743(4) & 0.6745(3) \\
       & $a^2E_2^{\rm kin}$ & 0.682(4) & 0.68(1) & 0.704(4) & 0.70(1) \\
       & $a^2E_1^{\rm spin}$ & -0.0127(1) & -0.0126(2) & -0.0129(1) & -0.0130(3) \\
       & $a^2E_2^{\rm spin}$ & -0.0115(9) & -0.011(1) & -0.011(1) & -0.012(1) \\\hline
6.4956 & $aE_1^{\rm stat}$ & 0.2742(3) & 0.275(1) & 0.2482(3) & 0.249(1) \\
       & $aE_2^{\rm stat}$ & 0.409(2) & 0.405(8) & 0.384(2) & 0.381(8) \\
       & $aE_3^{\rm stat}$ & 0.518(3) & 0.52(2) & 0.491(3) & 0.50(2) \\
       & $a^2E_1^{\rm kin}$ & 0.5999(5) & 0.5997(3) & 0.6240(3) & 0.6229(7) \\
       & $a^2E_2^{\rm kin}$ & 0.620(3) & 0.625(6) & 0.645(2) & 0.645(4) \\
       & $a^2E_1^{\rm spin}$ & -0.0081(1) & -0.0079(5) & -0.0076(1) & -0.0074(4) \\
       & $a^2E_2^{\rm spin}$ & -0.0108(9) & -0.005(4) & -0.0098(8) & -0.007(2) \\
\hline\hline
\end{tabular}
\end{center}
\caption{The measured values of the HQET energies in lattice units.
Shown are both the values obtained from a global fit (``Fit'') and from
our more conservative plateau selection (``Plateau'') as
described in the text, for both the HYP1 and HYP2 discretization of the
heavy quark.}
\label{tab:all_ens}
\end{table}
\begin{table}
\begin{center}
\begin{tabular}{llrrr|r}
\hline\hline
  & &  $\beta=6.0219$ & $\beta=6.2885$ & $\beta=6.4956$ & cont. limit\\\hline
$r_0\Delta E_{2,1}^{\rm stat}$ & HYP1 & 1.50(3) & 1.52(9) & 1.46(8) & 1.50(5) \\
                       & HYP2 & 1.51(3) & 1.51(3) & 1.47(8) & \\\hline
$r_0\Delta E_{3,1}^{\rm stat}$ & HYP1 & 2.67(10) & 2.78(9) & 2.7(2) & 2.7(1) \\
                       & HYP2 & 2.72(10) & 2.68(6) & 2.8(2) & \\\hline
$r_0\Delta E_{2,1}^{1/m}$ & HYP1 & 0.20(2) & 0.14(4) & 0.17(5) & 0.03(6) \\
                       & HYP2 & 0.21(2) & 0.08(4) & 0.13(3) & \\\hline
\raisebox{-3pt}{$r_0\Delta E_{2,1}^{\rm HQET}$} & HYP1 & 1.70(4) & 1.66(6) & 1.63(10) & 1.56(8)\\[-3pt]
                       & HYP2 & 1.72(4) & 1.59(6) & 1.59(9) & \\\hline
\multicolumn{3}{l}{$\left.r_0\Delta E_{2,1}^{\rm stat}\right|_{\rm continuum} +
 \left.r_0\Delta E_{2,1}^{1/m}\right|_{\rm continuum} $} &  &  & 1.54(9)\\[1.5ex]\hline
$r_0\Delta E_{\rm P-V}$ & HYP1 & -0.093(7) & -0.083(6) & -0.089(8) & -0.075(8) \\
                       & HYP2 & -0.114(9) & -0.103(8) & -0.092(7) & \\
\hline\hline
\end{tabular}
\end{center}
\caption{The energy level differences in units of the scale $r_0$.
Shown are the results at each $\beta$ for both static-quark actions, 
together with their common continuum limit.
\label{tab:en_diffs}}
\end{table}

\subsection{Continuum limit}

\begin{figure}[htb]
\begin{center}
\includegraphics[width=0.7\textwidth]{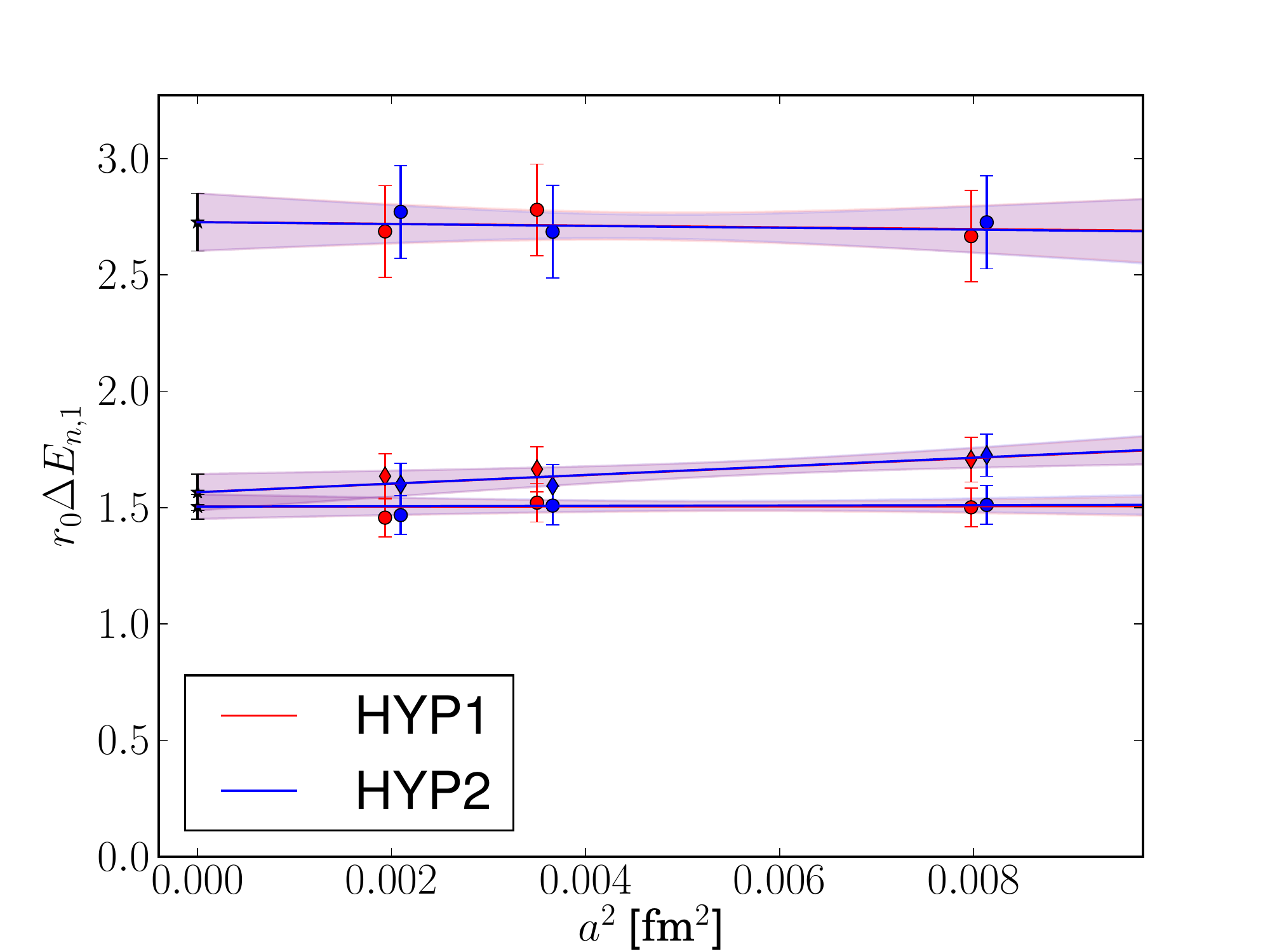}
\end{center}
\caption{Plot of the continuum limits (stars) of $\Delta E_{n,1}^{\rm stat}$ (circles) and $\Delta E_{2,1}^{\rm HQET}$ (diamonds). Shown are the results for both HYP1 (red, shifted to the left) and HYP2 (blue, shifted to the right).}
\label{fig:cl_spectrum}
\end{figure}

We now turn to the continuum extrapolation of the
level splittings. Using the fact that
the static actions employed are discretizations of the same continuum
theory, we perform a combined continuum limit by fitting a function of the
form ($k=1,2$ for HYP1, HYP2 actions)
\be
\Phi_{i,k}(a/r_0) = \Phi_i + c_{i,k} (a/r_0)^{s_i}
\ee
to our dimensionless quantitites 
$\Phi_i\in\{r_0\Delta E^{\rm stat}_{2,1},
r_0\Delta E^{\rm stat}_{3,1},  r_0\Delta E_{\rm P-V}, r_0\Delta E^{\rm HQET}_{2,1} \}$.
Since $\rmO(a)$-improvement is fully implemented in the static approximation,
we use powers $s_1=s_2=2$. On the other hand, the $\minv$ corrections
have $\rmO(a)$ discretization errors, yielding $s_3=1$ for the 
observable $\Delta E_{\rm P-V}$.
For $\Delta E_{n,1}^{\rm HQET}$, there are two possible ways of taking
the continuum limit. 
First we extrapolate 
$\Delta E_{n,1}^{\rm stat}$ and the $\rmO(\minv)$ contribution separately
to $a\to0$
and add the continuum limits afterwards. Here we set $s_4^\stat=2$ for the
static part and $s_4^{1/m}=1$ for the $\rmO(\minv)$ correction.
Second, one can form the combined
$\Delta E_{n,1}^{\rm HQET}$ at each $\beta$ and take the continuum limit
of the combination. The linear term in $a$ which is present
in the combined data, is suppressed by a factor $\minv$. 
Given in addition the flatness of the 
data in $a$ (see \fig{fig:cl_spectrum}) 
we just use $s_4=2$ for the combination, assuming that this
term dominates.

\begin{figure}[htb]
\begin{center}
\includegraphics[width=0.7\textwidth]{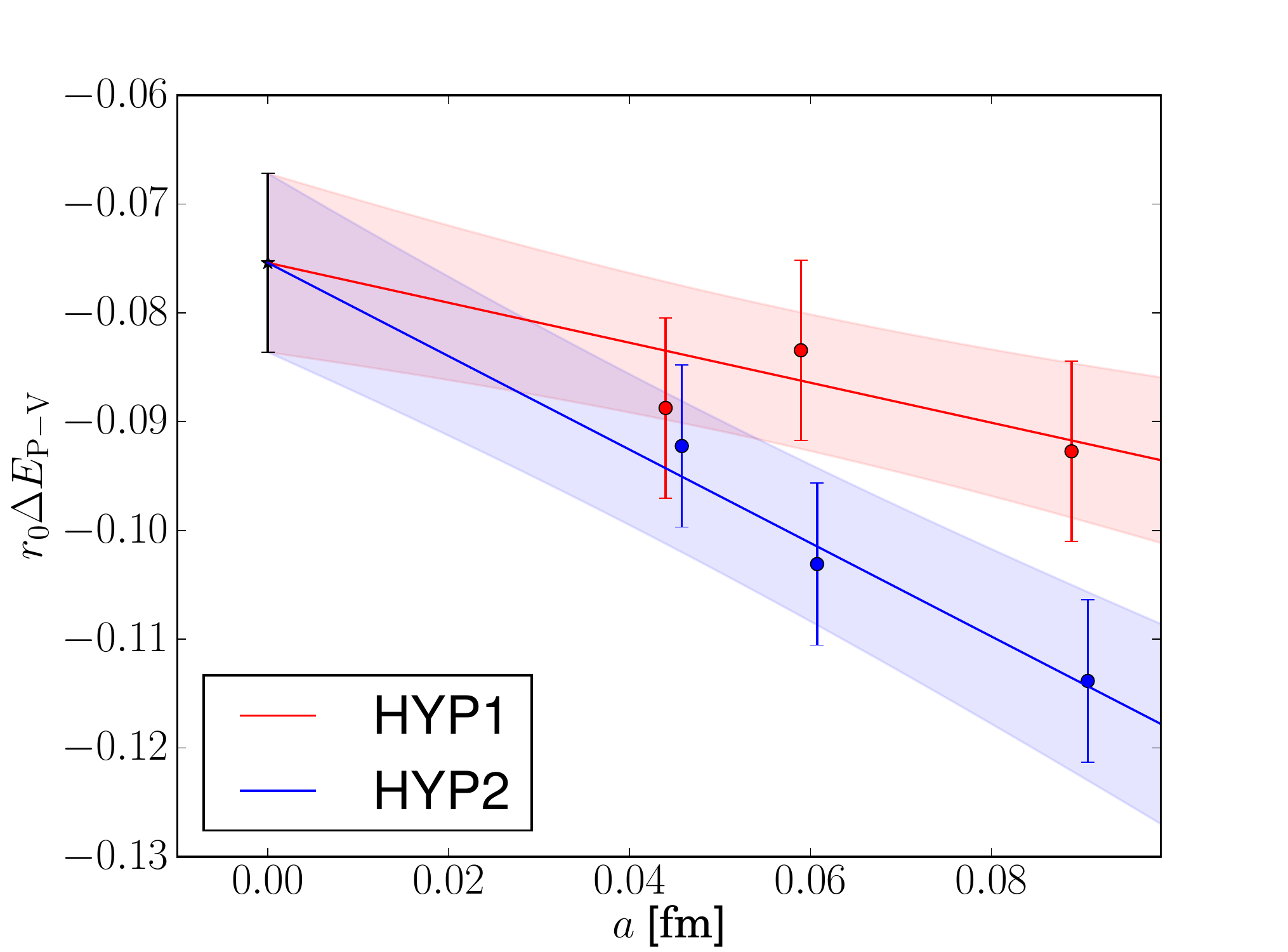}
\end{center}
\caption{Plot of the continuum limit (star) of $\Delta E_{\rm P-V}$ for HYP1 (red, shifted to the left) and HYP2 (blue, shifted to the right).}
\label{fig:cl_espin}
\end{figure}

\Fig{fig:cl_spectrum} shows the approach to the continuum limit
for the static and full HQET energy splittings. The results for HYP1 and
HYP2 are distinguished by color; the static splittings are shown as
circles, whereas the full HQET splitting $\Delta E_{2,1}^{\rm HQET}$ is
shown as diamonds. Also shown are the fits to the continuum limit together
with their error bands. We see that the approach to the continuum limit
is rather flat in particular for $\Delta E_{2,1}$, and that the $\rmO(\minv)$
corrections constitute only a small shift of the energy splitting
between the first excited and ground states.
In \fig{fig:cl_espin} we show the approach to the continuum limit
for the spin splitting $\Delta E_{\rm P-V}$ in the same fashion. 

\subsection{Results}

Our findings for the continuum values of the level splittings are summarized
in \tab{tab:en_diffs}. 

For the static spectrum, we obtain
\bea
r_0\Delta E_{2,1}^{\rm stat} &=& 1.50(5) \\
r_0\Delta E_{3,1}^{\rm stat} &=& 2.7(1) \,,
\eea
corresponding (using $r_0=0.5$ fm) to
$\Delta E_{2,1}^{\rm stat} = 594(21)$ MeV and 
$\Delta E_{3,1}^{\rm stat} = 1076(48)$ MeV
in good agreement with the results of
\cite{Burch:2008qx} at a fixed lattice spacing.
These numbers in physical units are meant as a rough illustration,
since no quenching error is attached to them. 

At $\rmO(\minv)$, we obtain $r_0\Delta E_{2,1}^{\rm HQET} = 1.54(9)$
when taking separate continuum limits for the static and $\rmO(\minv)$ energy
differences, and
\be
r_0\Delta E_{2,1}^{\rm HQET} = 1.56(8)
\ee
when combining static and $\rmO(\minv)$ energy levels before taking the
continuum limit. The results from both procedures agree within errors,
indicating that the $\rmO(a)$ term omitted in the combined extrapolation
is at most a minor source of systematic error.
In physical units (again using $r_0=0.5$ fm), our results correspond to 
$\Delta E_{2,1}^{\rm HQET} = 606(35)$ MeV and $617(31)$ MeV, respectively.

For the $B_s-B_s^*$ mass difference, we find
\be
r_0\Delta E_{\rm P-V} = -0.075(8) 
\ee
giving $\Delta E_{\rm P-V} = -29.8(3.2)$ MeV via $r_0=0.5$ fm. This is
to be contrasted to the experimental value
\cite{PDBook}
of $m_{B_s}-m_{B_s^*}=-49.0(1.5)$ MeV. We note that while $\rmO(\minv^2)$
effects may not be entirely negligible for such a small splitting, the
difference is too large to be entirely attributed to these.
Instead, a genuine quenching effect is involved.

The spin splitting between the first radial excitations in the pseudoscalar
and vector channels is
\be
r_0\Delta E_{\rm P'-V'} = -0.056(27)\,,
\ee
which is compatible with the spin splitting between the ground states.

Our results confirm the expectation that the $\rmO(\minv)$ contributions
are small compared to the static results.
To get an idea of the size of higher order corrections in $\minv$, we
have considered the dependence of the energy level splittings on the
matching conditions chosen in the determination
\cite{hqet:first1}
of the HQET parameters $\omegakin$ and $\omegaspin$.
We find that the dependence is very minor,
with a maximum deviation of $\delta|r_0 \Delta E_{\rm P-V}|=0.005$, less than
the statistical errors, and $\delta|r_0 \Delta E_{2,1}^{1/m}|=0.002$,
much less than the statistical errors. 
Comparing this to the naive power counting estimate of
$|O(\minv^2)| \sim (1/(r_0 \m_{\rm b})^2 \sim 1/100$, we see
that the tested $\minv^2$ terms are as small as expected or
even smaller. Note that $1/r_0 \approx 400\,\MeV$ is indeed
a typical non-perturbative QCD scale.

%% file: s4.tex
\section{Conclusions}
\label{s:conclusions}

In this paper, we have reported results from quenched lattice QCD
for the spectroscopy of the low-lying excited states of the $B_s$
and $B_s^*$ systems. An application of the generalized eigenvalue
method with all-to-all propagators 
to non-perturbative HQET at $\rmO(\minv)$ allows us to extract
precise results for the energies of the lowest-lying radial excitations
as well as for the $B_s-B_s^*$ splitting.
However, we emphasize again that a careful analysis of systematic errors
due to excited state contaminations is necessary. 

A first relevant observation to be pointed out concerns the renormalizability
of HQET. Unlike for QCD on and off the lattice,
there is no proof of renormalizability of the theory to all orders of
perturbation theory. However, we find that in our non-perturbative
computations  the divergences cancel after proper renormalization of HQET
\cite{hqet:first1}.
The left over lattice-spacing dependence in \fig{fig:cl_spectrum},
\fig{fig:cl_espin} is very flat. To appreciate this, note that 
\bes
   r_0^2 E_1^\mrm{kin} \approx (24\,,\;47\,,\;77) \mbox{ for } a=(0.1\,,\;0.08\,,\;0.05)\,\fm 
\ees
as seen in \tab{tab:all_ens} and weaker but still very prominent
divergences are present in $E_n^\mrm{stat}$. In other words we find strong
numerical evidence for the renormalizability of the theory; in fact also
the universality of the continuum limit is demonstrated in the figures.
It is also worth emphasizing that the present demonstration is
the first time the continuum limit is taken for mass splittings in HQET. 

We find the physical $\rmO(\minv)$ corrections to be small throughout.

The precision attained, in particular when taken together
with the relative smallness of the $\rmO(\minv)$ effects,
indicates that non-perturbative HQET combined with the use
of the GEVP for data analysis is a reliable  method for
determining $B$ meson spectra. We intend to apply it to the
$N_f=2$ case in the near future. In this context
one should remark that we were able to achieve good precision
using only 100 configurations in our quenched study.
Therefore we do expect to be able to
decrease the errors for dynamical fermions. However the influence
of topological modes being updated only slowly
\cite{lat09:stefan}
needs to be controlled or better algorithms with a faster 
decorrelation need to be used. A promising proposal has been made in
\cite{Luscher:2009eq}.

%% file: a1.tex
%%%%%%%%%%%%%%%%%%%%%%%%%%%%%%%%%%%%%%%%%%%%%%%%%%%%%%%%%%%%%%%%%%%%%%
\section{All-to-all propagators}\label{app:ata}
%%%%%%%%%%%%%%%%%%%%%%%%%%%%%%%%%%%%%%%%%%%%%%%%%%%%%%%%%%%%%%%%%%%%%%
\def\unity{{\mathbbm 1}} % needs package bbm
\def\vL{\hat{v}}         % low-mode vectors
\def\vR{\eta}            % random vectors
\def\dyadic{\cdot}
\def\NL{{N_L}}           % number of low modes
\def\NR{{N_\eta}}        % number of random vectors
\def\PL{{\mathbf P}_{\mathrm L}} %%% {{\mathbb P}_{\mathrm L}}
\def\PH{{\mathbf P}_{\mathrm H}} %%% {{\mathbb P}_{\mathrm H}}
\def\Pt{{\mathbf P}_{t}}

In this appendix, we explain the details of our implementation 
of all-to-all propagators, which follows the idea of
\cite{alltoall:dublin}
with some useful improvements.

\subsection{Even-odd preconditioning} 
%%%%%%%%%%%%%%%%%%%%%%%%%%%%%%%%%%%%%
To reduce the computational effort and storage requirement 
for the matrix inversions, we consider even-odd preconditioning
of the (hermitian) Wilson-Dirac operator $Q = 2\kappa \gamma_5 D$.
With even/odd ordering of the sites one has a block structure
\be\nonumber
   Q = \gamma_5 
       \left(\begin{array}{cc}
          M_{ee} & M_{eo} \\
          M_{oe} & M_{oo}
       \end{array}\right)
  \,,
\ee
where $M_{ee}$ ($M_{oo}$) differs from unity by the clover term on the even (odd) 
sites, and $M_{oe}$ ($M_{eo}$) is the hopping term. Defining
\be\nonumber
   B \equiv \left(\begin{array}{cc}
       \unity_e  & - M_{ee}^{-1}M_{eo} \\
       0         & \unity_o  \end{array}\right) 
   \,, 
   %%% \ \ 
   %%% B^\dagger = \gamma_5 \left(\begin{array}{cc}
   %%%     1                       &  0 \\
   %%%     - M_{oe} (M_{ee})^{-1}  & 1  \end{array}\right) \gamma_5,
\ee
the preconditioned matrix $B^\dagger Q B$ is block-diagonal 
and the propagator can be factorized as
\be
    Q^{-1} = B
             \left(\begin{array}{cc}
                 \hat{Q}_{ee}^{-1} & 0 \\
                 0            & \hat{Q}_{oo}^{-1} 
             \end{array}\right)
             B^\dagger
    \,,
    \label{invQ}
\ee
where $\hat{Q}_{ee} = \gamma_5 M_{ee}$ is diagonal in space-time, and
$\hat{Q}_{oo} = \gamma_5(M_{oo} - M_{oe} M_{ee}^{-1}M_{eo}) = \hat{Q}_{oo}^\dagger$.

\subsection{Approximate low modes and a stochastic estimator}
%%%%%%%%%%%%%%%%%%%%%%%%%%%%%%%%%%%%%%%%%%%%%%%%%%%%%%%%%%%%%
We consider an orthonormal basis  $\{ \vL_i: i = 1 \ldots \NL \}$ 
of an $\NL$ dimensional subspace ($\NL \ge 0$) of all fermion fields
which live only on odd sites. Defining the projectors
\be\nonumber
    \PL \equiv \sum_{i=1}^\NL \vL_i \dyadic \vL_i^\dagger
    \ \ {\rm and} \ \
    \PH \equiv \unity_{o} - \PL
    \,,
\ee
we can write 
\be
   \hat{Q}_{oo}^{-1} 
   = \hat{Q}_{oo}^{-1} (\PL+\PH)
   = \sum_{i=1}^\NL (\hat{Q}_{oo}^{-1} \vL_i) \dyadic {\vL_i}^\dagger + \hat{Q}_{oo}^{-1} \PH
   \,.
   \label{invQhatHL}
\ee
A natural choice for $\vL_i$ are approximate eigenvectors of the low-lying 
eigenvalues of $\hat{Q}_{oo}$
\be
   \hat{Q}_{oo} \vL_i = \lambda_i \vL_i + \hat{r}_i
    \,.
   \label{evQhat}
\ee
with $\|\vL_i\| = 1$ and ${\vL_i}^\dagger \hat{r}_k = 0$.
Then, the part $\hat{Q}_{oo}^{-1}\PL$ in (\ref{invQhatHL}) is expected to 
approximate the long-distance behaviour of the propagator \cite{alltoall:dublin,Neff:2001zr}, 
and the inversions\footnote{
    Since we do not explicitly use $\hat{Q}_{oo}^{-1} \vL_i \approx \lambda_i^{-1} \vL_i$, 
    the errors $\|\hat{r}_i\|$ in (\ref{evQhat}) are allowed to be large.
    In practice, we require $\|\hat{r}_i\|\le 0.001 \cdot \vert\lambda_i\vert$,
    and take $\lambda_i^{-1} \vL_i$ only as start vectors for the inversion.
} 
needed in $(\hat{Q}_{oo}^{-1}\vL_i)$ are cheap.

On the other hand, we can introduce a stochastic estimator for $\PH$. 
We take random vectors $\vR_i$ with
\bea
  \langle\vR_{i,\alpha}\rangle_\vR &=& 0\,, \label{eta1}\\
  \langle\vR_{i,\alpha}\,\vR_{j,\beta}^\ast\rangle_\vR &=& \delta_{ij} \delta_{\alpha\beta} \,, \\
  \langle\vR_{i,\alpha}\,\vR_{j,\beta}\rangle_\vR &=& 0 \,, \label{eta2}
\eea
where $\alpha, \beta$ denote combined (color, Dirac, and site) indices,
and $\langle . \rangle_\vR$ is the average over $\vR$.
The relations (\ref{eta1})--(\ref{eta2}) hold, for instance, in the case 
of a U(1) noise
\bea\nonumber
   \vR_{i,\alpha} = \exp(i\phi_{i,\alpha})\,,
\eea
where $\phi_{i,\alpha}$ are independently uniformly distributed in $[0,2\pi)$
(while for $Z_2$ noise the average in (\ref{eta2}) would not vanish for $i=j$). 
Thus, the second term in (\ref{invQhatHL}) can be written as
\be
   \hat{Q}_{oo}^{-1}\,\PH 
     = {1 \over \NR} \sum_{i=1}^\NR \big\langle  
             \hat{Q}_{oo}^{-1}\, \PH \, \vR_i \cdot \vR_i^\dagger 
       \big\rangle_{\vR}
   \,,
   \label{invQhatH}
\ee
and the estimator of $\hat{Q}_{oo}^{-1}$ can be written as a sum of dyadic products
\be
    \hat{Q}_{oo}^{-1} = \sum_{i=1}^{\NL+\NR} \big\langle \hat{w}_i \dyadic \hat{u}_i^\dagger \big\rangle_{\vR}
    \,,
    \label{invQhat}
\ee
with\footnote{
   One may also use
   $\hat{w}_i = \hat{Q}_{oo}^{-1}\,\hat{u}_i$ and 
   $\hat{u}_i = \NR^{-1/2}\,\PH\,\vR_i$
   for $i = \NL\!+\!1, \ldots, \NL\!+\!\NR$,
   but we have not tested this option.
}
\be\nonumber
   \begin{array}{rclrcll}
       \hat{w}_i & = & \hat{Q}_{oo}^{-1}\,\hat{u}_i \ , &
       \hat{u}_i & = & \vL_i &                      
       (i=1, \ldots, \NL) \\
       \hat{w}_i & = & \hat{Q}_{oo}^{-1}\,\PH\,\hat{u}_i \ , \ &
       \hat{u}_i & = & \NR^{-1/2}\,\vR_i &                      
       (i=\NL\!+\!1, \ldots, \NL\!+\!\NR) \\
   \end{array}
\ee
The full propagator $Q^{-1}$ is then obtained from (\ref{invQ}).
Since $B$ connects only adjacent time slices,  the block $\hat{Q}_{ee}^{-1}$ 
does not contribute to the propagator between sites with time
separation $\vert x_0-y_0\vert > 2a$. In this case, we can simply write
\be
    Q^{-1} = \sum_{i=1}^{\NL+\NR} \big\langle w_i \dyadic u^\dagger_i \big\rangle_{\vR}
    \label{invQdyadic}
    \,,
\ee
with 
\be\nonumber
  w_i \equiv B\left(\begin{array}{c}0\\\hat{w}_i\end{array}\right)
  \ \ {\rm and} \ \ 
  u_i \equiv B\left(\begin{array}{c}0\\\hat{u}_i\end{array}\right)
  \,.
  \label{WfromWhat}
\ee

Even-odd preconditioning can be seen as a form of dilution
\cite{alltoall:dublin},
since there are only half as many components of the noise field $\eta$
in the even-odd preconditioned case as without preconditioning. Note, however,
that unlike other dilution schemes, even-odd preconditioning does not
increase the number of inversions needed. 

\subsection{Time dilution}
%%%%%%%%%%%%%%%%%%%%%%%%%%
In addition, we use the more conventional time dilution scheme.
It is implemented by replacing $\PH$ in (\ref{invQhatHL}) by 
$\PH \sum_t \Pt$ where $\Pt$ projects on the components corresponding 
to (odd) sites with time coordinate $t$.
Then, an independent stochastic estimator is introduced for each term
\be\nonumber
  \PH \Pt = {1 \over \NR} \sum_{i=1}^{\NR}
            \big\langle (\PH\,\vR_{ti}) \dyadic \vR_{ti}^\dagger \big\rangle_{\vR}
  \,,
\ee
where the noise vectors $\vR_{ti}$ have non-vanishing components only for 
(odd) sites on time-slice $t$.

Note that due to the hopping term in $B$ the full propagator (\ref{invQdyadic})
from time slice $x_0$ to $y_0$ receives contributions 
from noise vectors $\vR_{ti}$ on three time slices, $t=x_0, x_0\pm a$,
i.e. three inversions are required for the propagator from one time slice $x_0$. 
However, a total of $T$ inversions is sufficient, and hence no 
extra effort is required, if one computes the propagator for 
all $x_0$, as we do in our measurements. 

Analysing the variance of a heavy-light two-point correlator
as described in \cite{Luscher:2010ae}, one sees that the
variance with time dilution decays roughly as $\rme^{-(x_0-y_0) m_\pi}$,
while the expression without time dilution contains
pieces independent of $x_0-y_0$. This renders time dilution very
profitable.